# *Generalisable prediction model of surgical case duration: multicentre development and temporal validation*

**Authors with institutional affiliations:**
Daijiro Kabata, PhD, MPH [a,*,†] (ORCID: 0000-0002-1243-4928)
Mari Ito, PhD [a,*,†] (ORCID: 0009-0000-6778-6763)
Tokito Koga, MD, MBA [b] (ORCID: 0009-0000-9423-872X)
Kazuma Yunoki, MD, MBA [c] (ORCID: 0000-0001-6684-8993)

[a] Center for Mathematical and Data Science, Kobe University, Kobe, JAPAN

[b] Department of Anesthesiology, Hyogo Prefectural Nishinomiya Hospital

[c] Department of Anesthesiology and Critical Care, Kobe City Medical Center General Hospital

[*] These authors contributed equally and are joint first and joint corresponding authors.

[†] Corresponding author 1: Daijiro Kabata, PhD, MPH, Center for Mathematical and Data Science, Kobe University, 1-1 Rokkodai-cho, Nada-ku, Kobe, Hyogo 657-8501, JAPAN (E-mail address: daijiro.kabata@port.kobe-u.ac.jp)

[†] Corresponding author 2: Mari Ito, PhD , Center for Mathematical and Data Science, Kobe University, 1-1 Rokkodai-cho, Nada-ku, Kobe, Hyogo 657-8501, JAPAN (E-mail address: mariito@opal.kobe-u.ac.jp)

# Abstract

**Background**

Accurate prediction of surgical case duration underpins operating room (OR) scheduling, yet existing models often depend on site- or surgeon-specific inputs and rarely undergo external validation, limiting generalisability.

**Methods**

We undertook a retrospective multicentre study using routinely collected perioperative data from two general hospitals in Japan (development: 1 January 2021–31 December 2023; temporal test: 1 January–31 December 2024). Elective weekday procedures with American Society of Anesthesiologists (ASA) Physical Status 1–4 were included. Pre-specified preoperative predictors comprised surgical context (year, month, weekday, scheduled duration, general anaesthesia indicator, body position) and patient factors (sex, age, body mass index, allergy, infection, comorbidity, ASA). Missing data were addressed by multiple imputation by chained equations. Four learners (elastic-net, generalised additive models, random forest, gradient-boosted trees) were tuned within internal–external cross-validation (IECV; leave-one-cluster-out by centre–year) and combined by stacked generalisation to predict log-transformed duration.

**Results**

We analysed 63,206 procedures (development 45,647; temporal test 17,559). Cluster-specific and pooled errors and calibrations from IECV are provided with consistent performance across centres and years. In the 2024 temporal test cohort, calibration was good (intercept 0.423, 95%CI 0.372 to 0.474; slope 0.921, 95%CI 0.911 to 0.932).

**Conclusions**

A stacked machine-learning model using only widely available preoperative variables achieved accurate, well-calibrated predictions in temporal external validation, supporting transportability across sites and over time. Such general-purpose tools may improve OR scheduling without relying on idiosyncratic inputs.

# Introduction

Efficient operation of hospital operating rooms (ORs) is important for patient flow, staff satisfaction, and financial sustainability of healthcare institutions [1, 2]. A key factor in OR efficiency is the accuracy of surgical case duration estimates used for scheduling [3]. Accurate predictions enable optimal OR scheduling, appropriate resource allocation, reduced patient waiting times, and minimal overtime for staff. However, traditional methods for estimating surgical time – such as a surgeon's subjective guess or simple averages from past cases – are often unreliable [1, 4]. Inaccurate time estimates can lead to OR over−utilisation or under−utilisation, with serious repercussions: overestimation leaves expensive OR resources idle, whereas underestimation causes surgeries to run late, delaying subsequent cases and increasing staff overtime [4, 5]. These inefficiencies ultimately drive up costs, contribute to case cancellations, and erode staff and patient satisfaction.

In recent years, artificial intelligence and machine learning (ML) techniques have emerged as promising tools to improve the precision of surgical duration forecasting. Modern ML algorithms can leverage large, complex perioperative datasets to capture patterns beyond the reach of human estimation. Numerous studies across different surgical specialties have reported that ML-based models can outperform traditional prediction methods in estimating case lengths [4, 6–8]. For example, data-driven models have achieved substantially lower prediction errors than surgeon estimates or simple historical averages, translating into more accurate scheduling and potential financial benefits for OR management [7]. These findings underscore the potential of ML to optimise OR utilisation by reducing scheduling errors [9]. Furthermore, prolonged operative durations are known to be associated with increased postoperative complications and infections [10, 11], so improvements in predicting and managing surgery times may also confer downstream benefits for patient outcomes.

Despite the promise of ML, existing surgical time prediction models face significant challenges that limit their practical utility and generalizability. First, many published ML models rely on predictors that are highly specific to a given institution or surgeon. For instance, some approaches incorporate the identity of the surgeon or use procedure descriptions in non-standard free-text format as key features [2, 6]. These idiosyncratic variables can indeed capture local practice patterns and improve predictive accuracy in the development setting, but they drastically reduce the model's reproducibility outside the original context. A model that heavily depends on a particular surgeon's past performance or on institution-specific coding of procedures will likely falter when applied to a different hospital or even as personnel and practices evolve over time. In other words, such models risk overfitting to their training environment, which undermines their usefulness for new patients or external institutions. Second, there is a pervasive lack of rigorous external validation in many ML-based OR time prediction studies. Most models are trained and tested only on single-centre data, calling into question their broader applicability. A recent systematic review of machine learning tools in perioperative medicine found that out of 103 studies evaluated, only 13% underwent any form of external validation, and roughly 75% were validated solely on data from the originating institution [12]. This paucity of external testing raises serious concerns about generalizability: a predictive algorithm that performs well on one hospital's data may not generalize to others with different patient populations, case mixes, or workflow patterns. As a result, many existing models remain confined to the setting in which they were developed and cannot be confidently deployed elsewhere. Indeed, without demonstrating robustness to external data, even the same institution may find that a model trained on last year's cases performs poorly on future cases as surgical techniques and team compositions change.

  In light of these gaps, our study aims to develop a more generalizable and reproducible model for predicting surgical case duration. We specifically address the limitations above by using only predictor variables that are widely available and not tied to a single surgeon or hospital. In constructing our model, we deliberately excluded features like surgeon identifiers or institution-specific procedure codes. Instead, we focused on general preoperative factors such as the patient's demographic and clinical factors and the general categories of surgical procedure. These predictors are commonly recorded across hospitals and are unlikely to vary in definition between institutions, which enhances the model's transportability. By restricting input variables to this generalizable set, we aim to create a robust prediction tool that maintains accuracy across different clinical environments and remains valid as conditions evolve over time. To further demonstrate its broad applicability, we trained and evaluated our model using a multicentre dataset from two general hospitals, including a held-out temporal test cohort to simulate prospective use. Finally, we implemented the finalized model as an interactive Shiny web application, providing a user-friendly interface for clinicians to input preoperative case information and obtain an immediate case duration prediction [13]. We designed and report this study in accordance with TRIPOD-AI and the TRIPOD-Cluster extension, explicitly accounting for clustered data and reporting cluster-specific performance in development and validation. In the following sections, we detail the development of this general-purpose prediction model and discuss its potential impact on improving OR scheduling and efficiency in diverse healthcare settings.

## Methods

### *Study Design and Setting*

We conducted a retrospective, multicentre observational study using routinely collected perioperative data from two general hospitals in the Kansai region of Japan: Kobe City Medical Center General Hospital and Hyogo Prefectural Nishinomiya Hospital. This study was designed and reported in accordance with the TRIPOD-AI and TRIPOD-Cluster guidelines [14, 15]. For TRIPOD-Cluster, we predefined clusters as centre–year units and applied internal–external cross-validation (IECV) via a leave-one-cluster-out (LOCO) procedure during model development and tuning, with cluster-specific and pooled performance reported. Hyogo Prefectural Nishinomiya Hospital contributed surgical cases performed from 1 January 2021 to 31 December 2024; Kobe City Medical Center General Hospital contributed cases from 1 January 2022 to 31 December 2024. Both institutions provide comprehensive surgical services across a broad range of specialties (for example, general surgery, orthopaedics, neurosurgery, urology, obstetrics and gynaecology, ophthalmology, otorhinolaryngology, cardiothoracic surgery, plastic surgery, and others).

The study protocol was approved by the institutional review board of Kobe City Medical Center General Hospital (approval number: zn251024) and Hyogo Prefectural Nishinomiya Hospital (approval number: R6-22). In line with national guidance for research using de-identified clinical records, we implemented an opt-out approach to informed consent and provided public notification.

### *Participants*

The sampling frame comprised 72,511 surgical records across both hospitals and years. We excluded emergency procedures to match the intended preoperative use case, leaving 65,529 records. Because weekend operating patterns differed across hospitals, we restricted the dataset to weekday procedures to ensure comparability. Moreover, to define a clinically homogeneous cohort and harmonise the distribution of baseline risk, we retained cases with an American Society of Anesthesiologists (ASA) Physical Status of 1 to 4. We then excluded records with a missing or implausible primary outcome (surgical case duration from room entry to room exit) for developing and validating the prediction model. This selection procedure produced a final analysis cohort of 63,206 procedures. A cohort-selection diagram is provided in Figure 1.

### *Outcome*

The primary outcome was surgical case duration in minutes, defined as the interval from operating-room entry to operating-room exit. In both hospitals, circulating nurses and clinicians enter time stamps into the theatre and anaesthesia information systems; durations are calculated automatically from these time stamps and verified during routine workflows. Because several candidate models assume approximately normal residuals and because surgical case duration is right-skewed, we analysed the natural logarithm of surgical case duration (log-transformed minutes).

### *Predictors*

To maximise reproducibility and transportability, we *a priori* limited predictors to variables that are widely available before or at the start of surgery in most hospitals. Surgical-context variables were the year, month, and weekday of surgery; scheduled duration (minutes) recorded in the scheduling system; an indicator for general anaesthesia (versus local anaesthesia) derived from structured anaesthesia fields; and intraoperative position encoded as separate indicators (supine, prone, sitting, lithotomy, lateral, other; multiple selections allowed per case where applicable). Patient variables were sex, age, body mass index, history of allergy, presence of infection, presence of comorbidity, and American Society of Anesthesiologists (ASA) Physical Status. We did not develop, tune, or report any models that incorporated surgeon identifiers or surgeon-level attributes, to maximise reproducibility and transportability.

### *Model Development, Tuning, and Stacking*

The development cohort comprised all eligible surgical procedures performed between 1 January 2021 and 31 December 2023, while the temporal test cohort comprised those performed between 1 January and 31 December 2024.

Missing predictor values were handled using multiple imputation by chained equations (MICE). The imputation models included all pre-specified predictors, a fixed effect for the centre-year cluster (the unit used for IECV), and the log-transformed outcome as an auxiliary variable to improve the plausibility of the missing-at-random assumption; the outcome itself was never imputed. For continuous variables we used predictive mean matching; for binary variables, logistic regression; and for unordered categorical variables, polytomous logistic regression. We generated five imputed datasets with five iterations per dataset. To avoid information leakage, imputation was performed within each IECV split using only the in-fold training data and then applied to the corresponding validation fold. All model fitting and hyper-parameter tuning were conducted independently within each imputed dataset, and estimates were combined across imputations using Rubin's rules. Model fitting, hyper-parameter tuning, model stacking, and performance estimation were repeated within each imputed dataset. Within each imputed development dataset, four prespecified base learners were trained to predict the log-transformed surgical case duration: random forest regression, generalised additive models, elastic-net regularised regression, and gradient-boosted trees. In line with TRIPOD-Cluster, we present cluster-specific metrics alongside pooled estimates from IECV [14, 16]. By validating across multiple centres and calendar years, this approach enables hyper-parameter estimation that is robust to both locational and temporal variation. For each learner, the hyper-parameters that minimised the mean squared error (MSE) across folds were selected, and the model was then refitted on the entire development dataset within each imputation to produce tuned base models for subsequent stack construction. The overall modelling workflow is summarised in Figure 2.

To improve predictive accuracy and stability, the tuned learners were combined using stacked generalisation. For each imputed dataset, out-of-fold predictions from the IECV were used to estimate weights for a convex combination of the learners; these weights were chosen to minimise the MSE on the log scale. The tuned base learners and the resulting stack were then refitted on the complete development dataset within each imputation to create a locked model.

The locked stack from each imputed dataset was applied to the temporal test cohort to obtain predictions. Calibration-in-the-large (intercept) and calibration slope were assessed by regressing the observed log-transformed surgical case duration on the predicted log-transformed surgical case duration, with an ideal intercept of 0 and slope of 1. The adjusted coefficient of determination ($R^2$) on the log scale was reported as a measure of explained variation in the temporal test cohort. Where appropriate, estimates were pooled across the five imputations using Rubin's rules.

All analyses were performed in R. Model training and resampling used caret with parallel back-ends; generalised additive models used mgcv; random forests used ranger; elastic-net models used glmnet; gradient-boosted trees used xgboost; and multiple imputation used mice. Reproducible scripts are available upon request and will be shared on acceptance.

*Deployment Note*

To facilitate reproducibility and potential clinical translation, we implemented an interactive Shiny application that returns the model-based prediction for surgical case duration when a user inputs the same predictors specified above. The app runs the locked stack created from the development data and is intended for research use only. The application is accessible at:

https://kabajiro.github.io/20250823_Surgical_Case_Duration_Calculator/ (temporal password: predict2025). In the deployed Shiny application, any missing user input fields are automatically imputed using the same multiple imputation by chained equations algorithm that was applied during model training. This ensures that the prediction is generated even if some inputs are incomplete, maintaining consistency with the model's development process.

# Results

*Participant Characteristics*

We analysed 63,206 elective weekday procedures (development data 45,647; temporal test data 17,559). Most cases were performed at Kobe City Medical Center General Hospital (overall 76.9%, 48,586/63,206; development 70.8%; test 92.7%), with the remainder at Hyogo Prefectural Nishinomiya Hospital (overall 23.1%, 14,620/63,206). By design, the temporal test cohort comprised only 2024 procedures, whereas the development cohort comprised 2021–2023 procedures.

Across months and weekdays, case volumes were broadly distributed (for example, by month ~7.4–10.1% in development; weekday proportions Monday–Friday 17.2–22.1% in test). Among records with anaesthesia mode available, general anaesthesia was used in 96.2%; body position at surgery (available for approximately one-third of cases) was most commonly supine (75.5% among recorded), followed by lithotomy (13.1%) and lateral (11.0%). The median scheduled duration was 150 minutes (first to third quartile 75–300), and the median observed duration from operating room entry to exit was 127 minutes (68–234). The sex distribution was balanced (female 49.3%), the median age was 69 years (52–78), and the median body-mass index was 21.95 kg/m² (18.83–24.75). A history of allergy was documented in 21.9%, infection indicators in 6.5%, and comorbidity in 20.2%. Among cases with an American Society of Anesthesiologists Physical Status recorded, the distribution was 1 17.5%, 2 49.0%, 3 31.9%, and 4 1.6%. Surgeon-level attributes were available for a subset (approximately one-half missing across fields); among recorded values, 29.5% held a doctoral degree, 52.2% were board-certified, 75.9% were listed as specialists, 45.3% had an instructing role, and 66.5% were male (Table 1).

### *Prediction Performance in IECV and Temporal External Validation*

Cluster-specific and pooled RMSE/MAE from the IECV are summarised in Table 2 and displayed in Figure 3. The error distributions were similar each other between the clusters. Applying the locked stack to the 2024 temporal test cohort, cluster-specific and pooled root mean squared error (RMSE) and mean absolute error (MAE) are summarised in Table 2 and displayed in Figure 3, and calibration is visualised in Figure 4. The calibration intercept was 0.423 (95% CI 0.372 to 0.474), and the calibration slope was 0.921 (95% CI 0.911 to 0.932), indicating negligible systematic bias and near-proportional agreement between predicted and observed log-transformed case times.

## Discussion

Our study demonstrates that a machine learning stacked model, trained on two general hospitals' data with only widely available preoperative variables, can accurately predict surgical case durations with high precision and reliability. In a held-out 2024 temporal test set, the model achieved low prediction errors as reflected by RMSE and MAE in IECV, with visually acceptable calibration (Figure 4). These results confirm that a robust, general-purpose prediction model can approach the accuracy of models that use institution-specific or surgeon-specific inputs, validating the viability of a transportable approach to OR scheduling.

      Our findings align with previous reports that modern ML methods outperform traditional estimates in surgical timing [8]. Prior studies across various specialties have reported substantial reductions in prediction error when using ML. For example, in an otolaryngology cohort, gradient-boosted trees improved mean absolute error (MAE) by ~8–10 minutes compared to standard methods [7]. Similarly, Babayoff *et al.* achieved an overall MAE of ~15 minutes using an stacked model on a single-hospital dataset [4]. Our stack's accuracy is comparable to these benchmarks, with prediction errors small based on the cluster-specific and pooled RMSE/MAE (Table 2), without relying on surgeon-specific inputs. Notably, unlike many earlier models, we achieved this performance *without* relying on surgeon identifiers or free-text procedure descriptions. This suggests that models built on generalizable features can attain near state-of-the-art

accuracy, challenging the notion that one must include idiosyncratic, site-specific data to get good predictions. Indeed, a recent multi-department study reported very high accuracy by tailoring separate models to each surgical specialty and including features like surgeon ID [17]. While that approach underscores how local customization can boost performance, those highly specific features also limit broader applicability. In contrast, our single model maintained excellent accuracy across two hospitals and diverse specialties *without* such customization, highlighting its potential generalizability.

Another strength of our study is the rigorous validation strategy used. We adopted an IECV framework during development, intentionally leaving out data by site and year to tune hyper-parameters in a way that is robust to institutional and temporal variation. This design minimizes overfitting to any one hospital's practices. We then confirmed performance prospectively on an external temporal cohort (all 2024 cases), effectively simulating a future implementation. The consistency of model accuracy and calibration in this temporal test suggests true reproducibility over time, which is critical for clinical usability. Few published models in this domain have undergone such strict validation. In fact, a recent systematic review of perioperative ML tools found that only ~13% of studies included any external validation across multiple centres [12, 18]. Our approach, therefore, represents a more reliable blueprint for prediction model development. We have shown that with careful design and validation, an OR time forecasting model can retain its performance when faced with new data – a property often lacking in prior single-centre studies [12]. Notably, we deliberately avoided surgeon- or site-specific inputs and still achieved excellent calibration and high explained variation in temporal external validation, supporting transportability across centres and over time in line with TRIPOD-Cluster recommendations. This finding is encouraging: it implies that widely available variables carry sufficient signal for accurate predictions, and that the model's generalizability is not achieved at the cost of predictive power. Hospitals without detailed surgeon data can still benefit from such a tool.

Despite these strengths, there is a limitation to consider. Our model was developed and evaluated using data from two large general hospitals in the Kansai region of Japan. While we deliberately restricted predictors to commonly recorded variables to enhance transportability, the model's generalizability to different healthcare systems or regions remains unproven. Practice patterns, patient demographics, and surgical workflows can vary internationally; thus, performance may differ in, for example, smaller community hospitals or non-Japanese populations. Validating the model on truly external datasets (in other countries or hospital networks) is a necessary next step.

In summary, we have developed a generalizable, high-performing ML model for surgical case duration prediction and validated it on temporal external data. By focusing on universally available preoperative inputs, we created a predictive tool that is not tied to a single surgeon or centre, making it suitable for diverse clinical settings. Our approach addresses the common reproducibility gaps in prior studies and offers a template for building prediction models that clinicians can trust across different hospitals and timeframes. Wider implementation of such models has the potential to markedly improve OR scheduling efficiency, reduce wasted resources, and ultimately support better care delivery in surgery [8].

## Author Contributions

Study concept and design: Kabata and Ito.

Acquisition, analysis, or interpretation of data: All authors.

Drafting of the manuscript: Kabata and Ito.

Critical revision of the manuscript for important intellectual content: All authors.

Statistical analysis: Kabata.

Obtained funding: Ito.

Administrative, technical, or material support: Kabata and Ito.

Study supervision: Kabata and Ito.

## Conflict of Interest

The authors declare no conflicts of interest related to this work.

## Financial Support

This work was partially supported by the Japan Society for the Promotion of Science (JSPS KAKENHI, grant number JP24K07946).

# Tables

## Table 1. Baseline demographic and clinical characteristics

| Characteristic | Overall<br>N = 63,206 | Development Data<br>N = 45,647 | Validation Data<br>N = 17,559 |
|---|---|---|---|
| Year of Surgery | | | |
| 2021 | 5.4% (3,416.0) | 7.5% (3,416.0) | 0.0% (0.0) |
| 2022 | 32.8% (20,727.0) | 45.4% (20,727.0) | 0.0% (0.0) |
| 2023 | 34.0% (21,504.0) | 47.1% (21,504.0) | 0.0% (0.0) |
| 2024 | 27.8% (17,559.0) | 0.0% (0.0) | 100.0% (17,559.0) |
| Month of Surgery | | | |
| 1 | 8.1% (5,102.0) | 7.4% (3,371.0) | 9.9% (1,731.0) |
| 2 | 7.9% (4,969.0) | 7.1% (3,225.0) | 9.9% (1,744.0) |
| 3 | 8.7% (5,489.0) | 8.1% (3,715.0) | 10.1% (1,774.0) |
| 4 | 8.5% (5,395.0) | 8.8% (4,003.0) | 7.9% (1,392.0) |
| 5 | 8.4% (5,283.0) | 8.4% (3,844.0) | 8.2% (1,439.0) |
| 6 | 8.5% (5,350.0) | 8.8% (4,025.0) | 7.5% (1,325.0) |
| 7 | 8.1% (5,127.0) | 8.2% (3,739.0) | 7.9% (1,388.0) |
| 8 | 8.2% (5,195.0) | 8.5% (3,883.0) | 7.5% (1,312.0) |
| 9 | 8.0% (5,031.0) | 8.4% (3,816.0) | 6.9% (1,215.0) |
| 10 | 8.8% (5,580.0) | 9.0% (4,088.0) | 8.5% (1,492.0) |
| 11 | 8.5% (5,397.0) | 8.8% (4,015.0) | 7.9% (1,382.0) |
| 12 | 8.4% (5,288.0) | 8.6% (3,923.0) | 7.8% (1,365.0) |
| Weekday of Surgery | | | |
| Monday | 20.5% (12,947.0) | 21.5% (9,803.0) | 17.9% (3,144.0) |
| Tuesday | 21.2% (13,396.0) | 21.3% (9,701.0) | 21.0% (3,695.0) |
| Wednesday | 19.0% (12,033.0) | 18.5% (8,440.0) | 20.5% (3,593.0) |
| Thursday | 21.7% (13,742.0) | 21.6% (9,870.0) | 22.1% (3,872.0) |
| Friday | 17.5% (11,088.0) | 17.2% (7,833.0) | 18.5% (3,255.0) |
| Admission | 50.3% (31,780.0) | 51.1% (23,344.0) | 48.0% (8,436.0) |
| General anaesthesia | 96.2% (21,794.0) | 96.1% (15,688.0) | 96.6% (6,106.0) |
| Missing % | 64% | 64% | 64% |
| Surgical position | | | |
| Supine Position | 75.5% (17,206.0) | 76.7% (12,628.0) | 72.3% (4,578.0) |
| Prone Position | 4.6% (1,053.0) | 4.6% (759.0) | 4.6% (294.0) |
| Sitting Position | 1.0% (233.0) | 1.2% (193.0) | 0.6% (40.0) |
| Lithotomy Position | 13.1% (2,982.0) | 13.2% (2,182.0) | 12.6% (800.0) |
| Lateral Position | 11.0% (2,503.0) | 10.4% (1,716.0) | 12.4% (787.0) |
| Other Position | 0.9% (213.0) | 0.9% (150.0) | 1.0% (63.0) |
| Missing % | 64% | 64% | 64% |
| Scheduled Surgery Duration (min) | 150.00 (75.00, 300.00) | 150.00 (75.00, 300.00) | 150.00 (90.00, 240.00) |
| Missing % | 32% | 30% | 38% |
| Actual Surgery Duration (minutes) | 127.00 (68.00, 234.00) | 127.00 (66.00, 244.00) | 126.00 (73.00, 215.00) |
| Sex: Female | 49.3% (31,187.0) | 49.9% (22,774.0) | 47.9% (8,413.0) |
| Age (years) | 69.00 (52.00, 78.00) | 70.00 (53.00, 78.00) | 68.00 (52.00, 77.00) |
| Body Mass Index | 21.95 (18.83, 24.75) | 22.01 (19.02, 24.76) | 21.77 (18.35, 24.72) |
| Missing % | 2.40% | 3.10% | 0.70% |
| History of Allergy | 21.9% (13,843.0) | 24.5% (11,201.0) | 15.0% (2,642.0) |
| Presence of Infection | 6.5% (4,115.0) | 6.4% (2,938.0) | 6.7% (1,177.0) |
| Comorbidity | 20.2% (12,787.0) | 18.3% (8,346.0) | 25.3% (4,441.0) |
| ASA Physical Status Classification | | | |
| 1 | 17.5% (3,976.0) | 17.6% (2,880.0) | 17.3% (1,096.0) |
| 2 | 49.0% (11,102.0) | 48.2% (7,890.0) | 50.8% (3,212.0) |
| 3 | 31.9% (7,241.0) | 32.4% (5,301.0) | 30.7% (1,940.0) |
| 4 | 1.6% (360.0) | 1.7% (283.0) | 1.2% (77.0) |
| Missing % | 64% | 64% | 64% |

Baseline demographic and clinical characteristics in the overall cohort (N = 63,206), the development data (N = 45,647), and the temporal test data (N = 17,559). Values are n (%) for categorical variables and median (IQR) for continuous variables. Percentages are calculated among non-missing observations; the "Missing %" rows report the proportion missing among all records. Surgical position may exceed 100% because multiple positions could be recorded per case. Abbreviations: ASA, American Society of Anesthesiologists; IQR, interquartile range.

**Table 2. Error metrics via IECV**

| Cluster | N | RMSE | MAE | Calibration Intercept | Calibration Slope |
|---|---|---|---|---|---|
| Site 1 at 2021 | 3416 | 0.538 [0.522 - 0.555] | 0.372 [0.359 - 0.385] | -0.894 [-0.962 - -0.826] | 1.134 [1.121 - 1.147] |
| Site 1 at 2022 | 4830 | 0.561 [0.546 - 0.576] | 0.385 [0.374 - 0.397] | -0.942 [-1.002 - -0.881] | 1.158 [1.146 - 1.169] |
| Site 1 at 2023 | 5098 | 0.543 [0.529 - 0.557] | 0.367 [0.356 - 0.378] | -0.807 [-0.863 - -0.752] | 1.126 [1.115 - 1.137] |
| Site 2 at 2022 | 15897 | 0.570 [0.562 - 0.579] | 0.413 [0.407 - 0.419] | 0.888 [ 0.827 - 0.950] | 0.834 [0.821 - 0.847] |
| Site 2 at 2023 | 16406 | 0.589 [0.581 - 0.598] | 0.413 [0.407 - 0.419] | 1.094 [ 1.033 - 1.154] | 0.789 [0.776 - 0.802] |
| Overall | 45647 | 0.561 [0.543 - 0.580] | 0.391 [0.371 - 0.410] | -0.132 [-1.034 - 0.770] | 1.008 [0.850 - 1.166] |

Cluster-specific and pooled error metrics obtained via internal–external cross-validation (IECV) with clusters defined by centre–year. Site 1 and Site 2 are Hyogo Prefectural Nishinomiya Hospital and Kobe City Medical Center General Hospital, respectively. Root mean squared error (RMSE) and mean absolute error (MAE) are computed on the natural log scale of surgical case duration. Calibration intercept (ideal = 0) and slope (ideal = 1) are estimated by regressing observed on predicted log-duration within each held-out cluster; the "Overall" row summarises pooled performance across IECV folds.

# Figures

**Figure 1. Flow diagram for the selection of the analytical population**

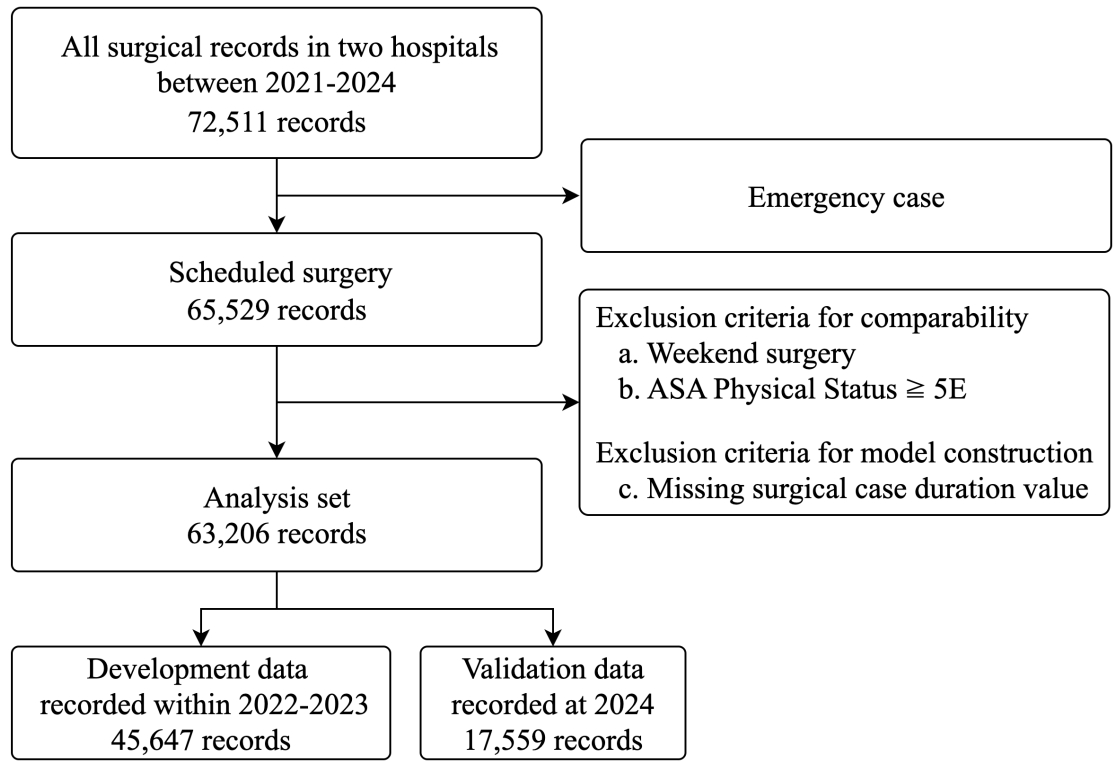

STROBE-style flow diagram showing cohort construction across 2021–2024 [19]. After excluding emergencies and weekend procedures and restricting to ASA 1–4 with plausible outcomes, the final analytic cohort was 63,206 procedures (development 45,647; temporal test 17,559).

**Figure 2. Flow diagram for the model construction with missing imputation**

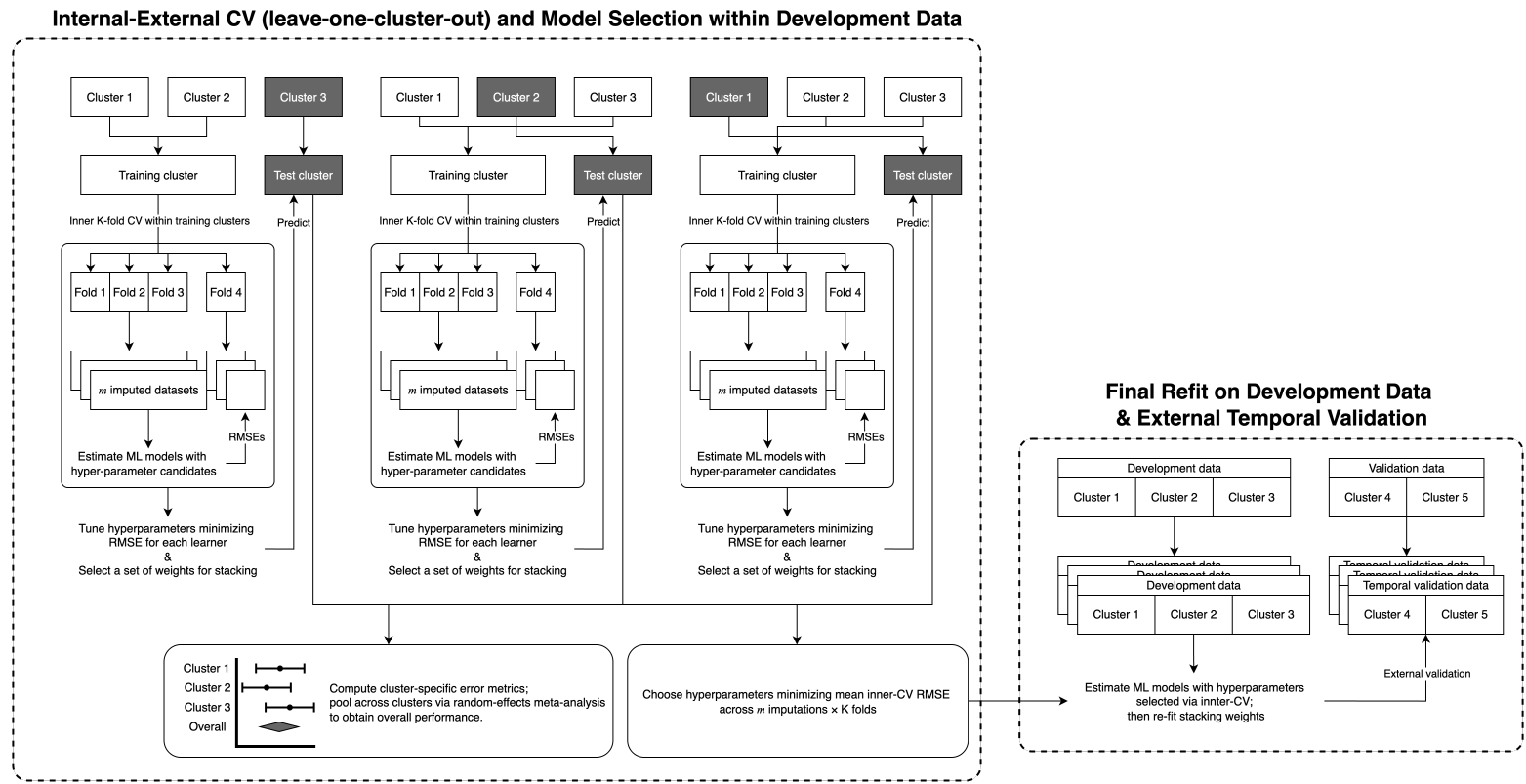

Overview of the modelling workflow. Within the development data, MICE addressed missing predictors; IECV (leave-one-cluster-out by centre–year) was used for training and tuning four pre-specified base learners (elastic-net, generalised additive model, random forest, gradient-boosted trees), followed by stacked ensembling to obtain a locked model for temporal testing. Imputation was performed within each IECV split to avoid information leakage; estimates were combined across imputations using Rubin's rules. This example including five clustered datasets employed K (inner-CV folds) = 4 and *m* (number of imputation) = 3.

**Figure 3. Cluster-specific and pooled RMSEs via IECV**

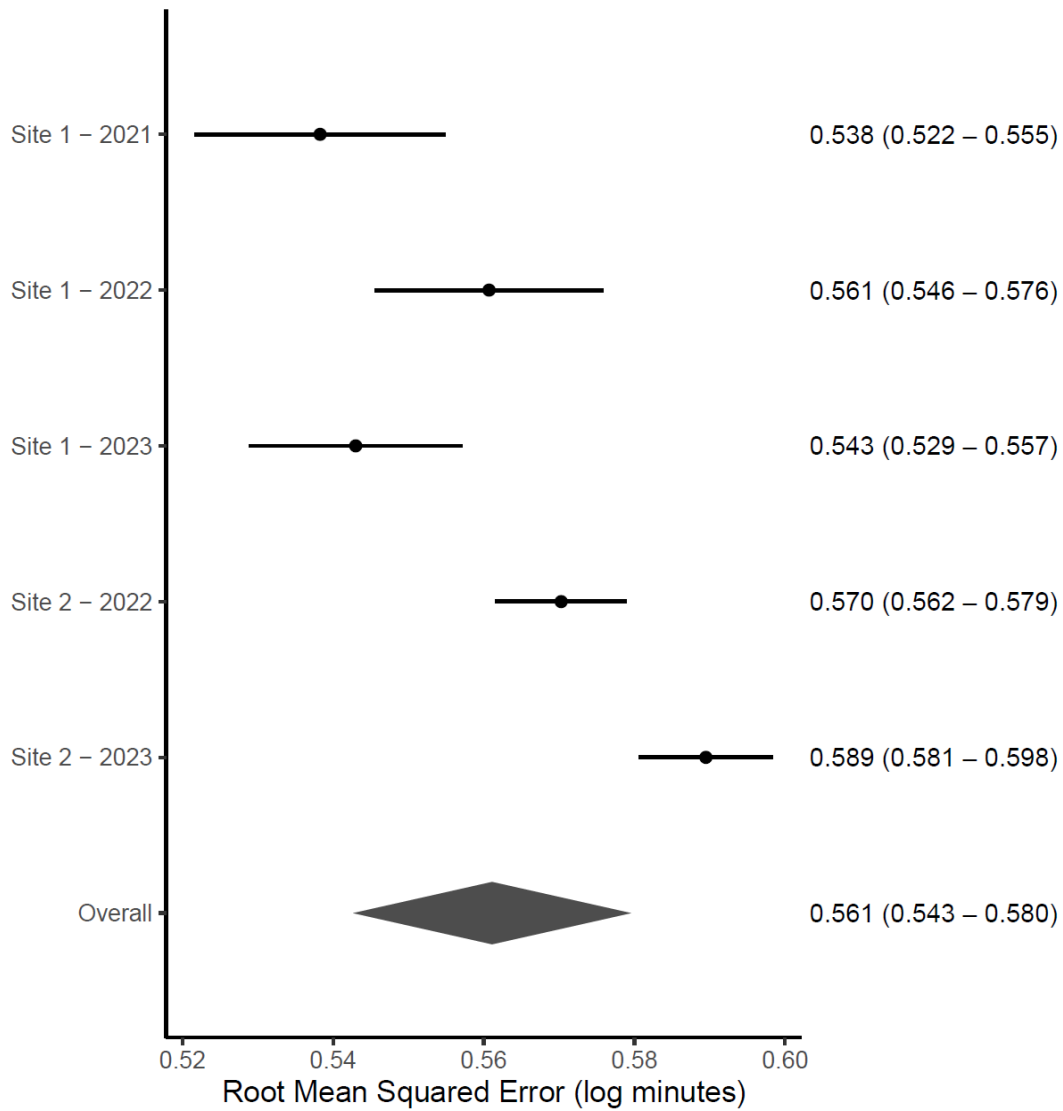

Cluster-specific and pooled prediction errors from IECV. Bars (or points) show RMSE for each held-out centre–year cluster with the pooled value summarised across folds; errors are evaluated on the log-transformed surgical duration. MAE is reported in Table 2. Site 1 and Site 2 are Hyogo Prefectural Nishinomiya Hospital and Kobe City Medical Center General Hospital, respectively.

**Figure 4. Calibration of predicted versus observed log-duration in the temporal test cohort**

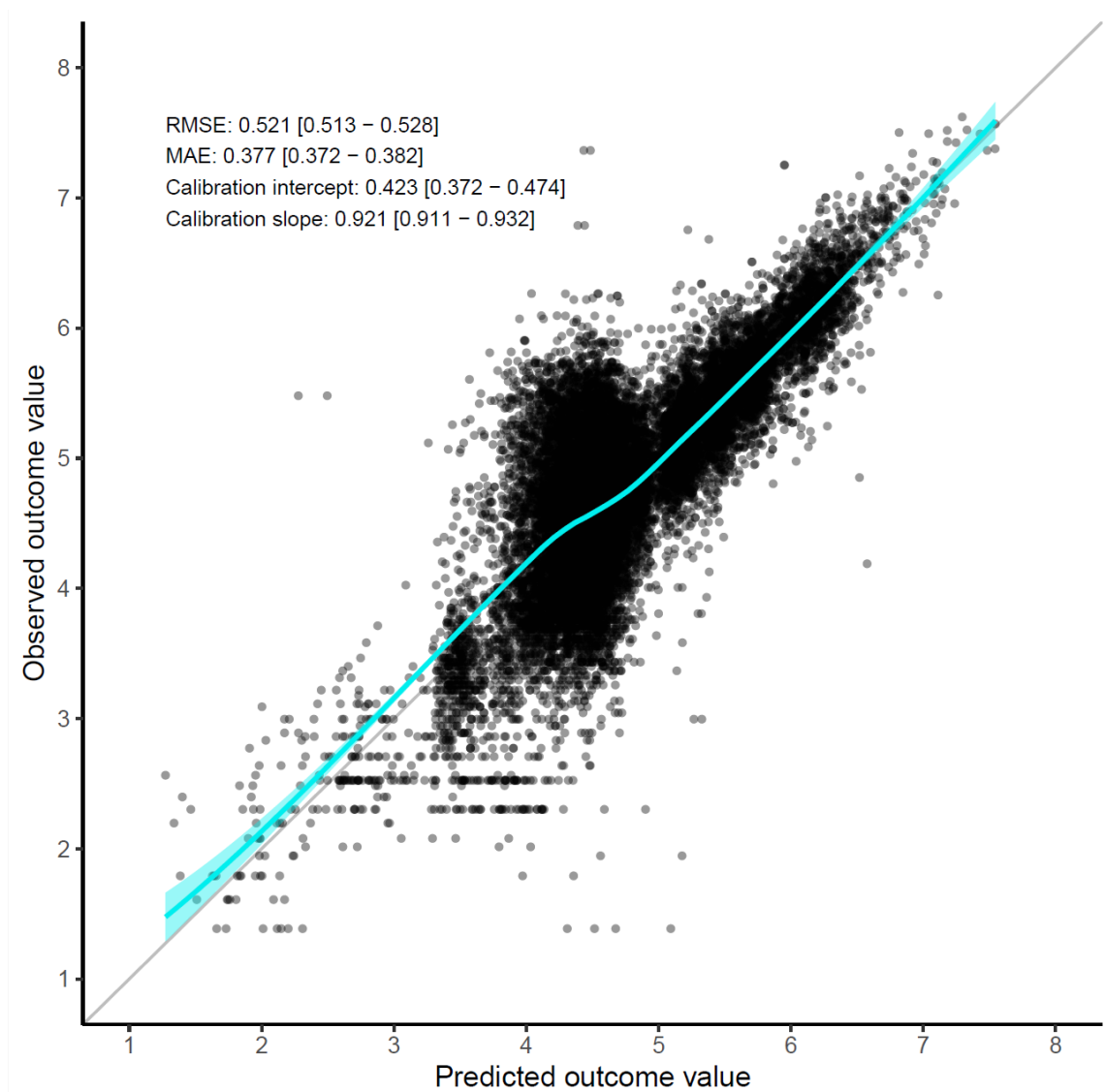

Calibration of predicted versus observed log-duration in the 2024 temporal test cohort. The 45-degree line indicates perfect calibration (intercept = 0, slope = 1). Points/curves depict agreement across the range of predictions; intercept and slope estimates are reported in Table 2 (development IECV) and summarised in Results for the temporal test.